\documentclass[11pt]{article}
\usepackage[T1]{fontenc}
\usepackage{lmodern}
\usepackage[a4paper,margin=25mm]{geometry}
\usepackage{microtype}
\usepackage{booktabs}
\usepackage{graphicx}
\usepackage{xcolor}
\usepackage{tikz}
\usetikzlibrary{arrows.meta,positioning,calc}
\usepackage{pgfplots}
\pgfplotsset{compat=1.18}
\usepgfplotslibrary{groupplots}
\usepackage{caption}
\usepackage{enumitem}
\usepackage{placeins}
\usepackage[numbers,sort&compress]{natbib}
\usepackage{hyperref}
\hypersetup{colorlinks=true,linkcolor=black,citecolor=black,urlcolor=black}

\definecolor{slotA}{HTML}{2A78D6}
\definecolor{slotB}{HTML}{EB6834}
\definecolor{slotC}{HTML}{1BAF7A}
\definecolor{slotD}{HTML}{EDA100}

\pgfplotsset{
  every axis/.append style={font=\sffamily\footnotesize, axis line style={black!45}, tick style={black!45},
    major grid style={black!10}, ymajorgrids, legend style={draw=none, fill=none, font=\sffamily\scriptsize},
    label style={text=black!80}, tick label style={text=black!70}},
}

\title{Lessons Outlive Their Failures: \\ A Case Study of Rules for AI Coding Agents}
\author{Hassan Hassanzadeh Aliabadi\\ Independent researcher\\ \texttt{h.h.aliabadi@gmail.com}}
\date{October 2026}

\begin{document}
\maketitle

\begin{abstract}
AI coding agents increasingly work inside a harness: rule files such as \texttt{AGENTS.md}, on-demand skills, runtime hooks and automated checks. Most studies of harnesses look at a single point in time or at short benchmark runs. I report a six-week case study of my personal system of record, a plain-markdown repository that three AI agent tools edited daily under one shared rule set. I built a ledger of 118 agent failures, their fixes and whether they came back. I also tested the system's monthly LLM documentation audit against 47 known stale statements and 20 planted errors. Finally, I replayed five past failures in isolated git clones, with and without their fixes. All results come from one repository with three runs per condition, so I report them as exploratory. Stale documentation (\emph{drift}) was the largest failure class (48 of 118), although 35 of those were found on one audit day. A single audit run found about half of the known drift (Claude Sonnet 5.5 mean: 50\,\%), and at least 93\,\% of its findings were correct. On Sonnet 5.5, the agent finished four of the five replayed tasks. In three of those four, the failure did not happen, with or without its fix. Only the stale-count failure came back without its check, and the check stopped it in every run. On older models, an agent drafted an em-dash once while the em-dash lint check was active, and the check rejected it. A written rule on run-time estimates did not change Sonnet 4.5's overestimates. These results suggest that harness lessons can outlive the failures they were written to fix, while drift persists and still needs mechanical checks. Tools that evolve a harness add lessons faster than they retire them, so they should re-test their rules from time to time and retire the obsolete ones.
\end{abstract}

\section{Introduction}

An AI coding agent does not act on its language model alone. It works inside a \emph{harness}: the instructions it loads at session start, the skills it loads on demand, the hooks that run around its tool calls, and the checks that run before a commit is accepted \cite{galster2026harness,weng2026harness,yang2024sweagent}. Rule files in the repository, such as \texttt{AGENTS.md} and \texttt{CLAUDE.md}, have become the standard way to give these instructions \cite{agentsmd2026,mohsenimofidi2026context}. Recent studies describe what these files contain \cite{chatlatanagulchai2025manifests,chatlatanagulchai2026readmes,jiang2026cursor}, how they grow over time \cite{chakrabarti2026growing,cai2026rules}, and how they affect task cost and completion \cite{lulla2026efficiency,gloaguen2026evaluating,zhang2026guardrails}.

Two gaps remain. First, almost all existing studies are cross-sectional: they mine snapshots of open-source repositories or measure single-turn benchmark runs. Galster et al.\ call for longitudinal and experimental studies of harnesses in daily use \cite{galster2026harness}. Second, people who maintain a harness respond to agent failures by adding \emph{lessons}. A lesson is either a prose rule that the model reads or an executable check that the runtime enforces. Recent frameworks automate this loop: Harness$^2$ keeps a growing library of lessons across tasks \cite{xu2026harnesssquared}, and Harness-R1 writes executable harness patches from failure traces \cite{shao2026harnessr1}. Yet little is known about what happens to a lesson after it enters the harness. Does the failure come back? Is the lesson still needed after the model is upgraded?

To answer these questions, I studied one harness in daily use for six weeks. The system is my personal system of record: one git repository of plain markdown files and Python scripts that holds my job search, my language learning and my life administration. Three agent runtimes (three different AI agent tools) edit this repository daily under one shared rule set. From the start of the study period, I enforced a logging policy: every repeatable agent failure and its fix had to be recorded with the reason. The commit history and change logs therefore record every harness change made through a commit or a log entry.

I investigate three research questions:
\begin{enumerate}[label=\textbf{RQ\arabic*},leftmargin=*,itemsep=2pt]
\item Which agent failures occur, which come back after a fix, and what side effects do the fixes have?
\item How much stale documentation (\emph{drift}) does an automated LLM audit find?
\item Does a past fix still change agent behaviour when the original failure is replayed, and does this depend on the model version?
\end{enumerate}

The paper makes three contributions. First, a ledger of 118 agent failures, coded in two LLM passes whose disagreements I resolved, with daily measurements of how much rule text the harness holds (Section~\ref{sec:method}). Second, a test of how much drift an LLM documentation audit finds, against an answer key of 47 real stale statements and 20 planted errors. Third, a replay of five past failures in throwaway git clones, with and without their fixes, on several model versions (a \emph{counterfactual replay}). The results (Section~\ref{sec:results}) point to one main conclusion: \emph{lessons can outlive their failures}. A behavioural failure is a mistake in what the agent does, such as quoting an outdated decision. On the current model, I could replay one behavioural failure with its written rule removed, and it did not recur. Two other failures did not occur even without their fixes. Documentation drift, by contrast, comes from how the repository changes, not from how the agent behaves. It came back again and again as the repository changed, a single LLM audit pass caught only half of it, and code checks were still needed to block it.

\section{Background and related work}
\label{sec:related}

\paragraph{Rule files and harness configuration.}
Mining studies of \texttt{CLAUDE.md}, \texttt{AGENTS.md} and Cursor rules show that these files mostly specify build commands, coding conventions and architecture summaries, with relatively few security or safety rules \cite{chatlatanagulchai2025manifests,chatlatanagulchai2026readmes,jiang2026cursor}. Galster et al.\ identify eight configuration mechanisms across agentic coding tools. They find that static context files dominate practice, while modular skills and subagents are underused \cite{galster2026harness}. Context files grow through small additions \cite{chatlatanagulchai2026readmes}. Across 1{,}867 repositories, Chakrabarti finds that a file grows by 226\,\% on average over its lifetime, and that older instructions are less likely to be deleted \cite{chakrabarti2026growing}. In a survey of 99 practitioners, Cai et al.\ found that 78\,\% of developers reported modifying rules to correct AI errors, most often by appending prohibitions \cite{cai2026rules}. Dos Santos et al.\ catalogue the resulting configuration smells, such as rule bloat and contradictory directives \cite{santos2026smells}. Vendor guides stress that prompt files give advisory context, not guaranteed enforcement. They recommend runtime hooks for hard constraints \cite{anthropic2026memorydocs,anthropic2026hooksdocs} and on-demand skills to keep always-loaded context small \cite{rajasekaran2025context,zhang2025skillsblog,xu2026agentskills}. Yet a factorial experiment across 1{,}650 sessions found no statistically significant effect of context file length, ordering or conflicting instructions on how well agents follow instructions \cite{mcmillan2026adherence}.

\paragraph{Do rules work?}
Evidence on the effectiveness of prose rules is mixed. Lulla et al.\ report reduced execution time and token consumption when an \texttt{AGENTS.md} file is present \cite{lulla2026efficiency}. Conversely, Gloaguen et al.\ find that repository context files do not consistently improve task resolution rates while increasing token costs by more than 20\,\% \cite{gloaguen2026evaluating}. Zhang et al.\ observe that the only standalone rules that improved performance were explicit prohibitions, and that randomly sampled rules performed comparably to human-curated ones \cite{zhang2026guardrails}. Converting natural-language instructions into executable verification checks increases compliance from 67.0\,\% to 88.3\,\% in ContextCov \cite{sharma2026contextcov}, while runtime policy enforcement in AgentSpec prevents the vast majority of unsafe tool invocations \cite{wang2025agentspec}. My counterfactual replay examines this same contrast between prose rules and executable code checks inside a lived production repository.

\paragraph{Automated harness evolution.}
Several recent systems automate the refinement of harnesses. Harness$^2$ iteratively adapts harness configurations per task and accumulates a persistent library of lessons for future runs \cite{xu2026harnesssquared}. Lin et al.\ demonstrate in Agentic Harness Engineering that performance gains stem primarily from tool design, middleware, and memory rather than system-prompt edits \cite{lin2026ahe}. Kang et al.\ identify \emph{harness-level forgetting}, in which a new harness modification degrades previously working behaviour, and guard each update with a regression check \cite{kang2026harnesscl}. Wang et al.\ further show that automated harness evolution frequently overfits to training benchmarks and fails to outperform simple baselines on held-out tasks \cite{wang2026rethinkingharness}. Existing automated harness systems add lessons or patches and check them against earlier behaviour under a fixed model. To my knowledge, none evaluates whether accumulated lessons remain necessary after the underlying language model is updated.

\paragraph{Agent memory and documentation staleness.}
Agent memory architectures store past episodes as retrievable text \cite{packer2024memgpt,park2023generativeagents,sumers2024coala,hu2026memorysurvey}, yet Chao et al.\ demonstrate that LLM agents struggle to recognise when stored facts have become outdated \cite{chao2026stale}. In software engineering, detecting outdated code element references \cite{tan2022outdatedrefs}, stale comments \cite{huang2024outdatedcomments}, and code-documentation inconsistencies \cite{kiecker2026cascade} is a long-standing problem. Because an agent treats its harness and repository documentation as ground-truth instructions, documentation drift can directly induce agent errors.

\paragraph{Agent failures and the closest study.}
Prior work has established failure taxonomies for multi-agent pipelines \cite{cemri2025mast} and evaluated destructive tool actions or false completion reports in red-team sandboxes \cite{shapira2026chaos,ruan2024toolemu,zhang2025agentsafetybench}. Interactive permission prompts carry their own human factors cost: users approve 93\,\% of prompts, and automated classifiers still miss some over-eager edits \cite{hughes2026automode}. Field studies indicate that coding agents increase short-term commit velocity \cite{he2026cursor,murphyhill2026cliagents}, while Cursor adoption also brings persistent increases in static-analysis warnings and code complexity \cite{he2026cursor}. The closest study to mine is Vasilopoulos \cite{vasilopoulos2026codified}, who describes building a codified context infrastructure with an always-loaded constitution, retrieval hooks, and on-demand specifications across 283 sessions in a single-developer codebase. Vasilopoulos reports that this infrastructure prevented repeated failures in four case studies. The present study differs by tracking failure recurrence by fix type, evaluating automated drift detection against a ground-truth answer key, and replaying historical failures across model versions.

\section{The system I studied}
\label{sec:setting}

\begin{figure}[!t]
\centering
\resizebox{\linewidth}{!}{\begin{tikzpicture}[
  font=\sffamily\footnotesize,
  stage/.style={rounded corners=2.5pt, align=center, minimum width=24.5mm, minimum height=17.5mm, text width=23mm, inner sep=2.5pt},
  text-rule/.style={stage, fill=slotA!13, draw=slotA!85, line width=0.8pt},
  code-check/.style={stage, fill=slotB!13, draw=slotB!85, line width=0.8pt},
  model-audit/.style={stage, fill=slotA!13, draw=slotB!85, line width=1.1pt, densely dashed},
  runtime/.style={draw=black!55, rounded corners=2.5pt, fill=white, align=center, minimum width=38mm, minimum height=9.5mm, inner sep=3pt, line width=0.6pt},
  repo/.style={draw=black!65, rounded corners=3pt, fill=black!4, align=center, minimum width=165mm, minimum height=8.5mm, line width=0.7pt},
  loopbox/.style={draw=black!60, rounded corners=2.5pt, fill=white, align=center, minimum height=9.5mm, inner sep=3pt, line width=0.6pt},
  arr/.style={-{Stealth[length=2.2mm, width=1.6mm]}, black!65, line width=0.75pt},
  lbl/.style={font=\sffamily\scriptsize, text=black!75},
]

\node[anchor=west, font=\sffamily\scriptsize\bfseries, text=black!80] at (-1.4, 4.15) {Legend:};
\node[draw=slotA!85, fill=slotA!13, rounded corners=1.5pt, line width=0.8pt, minimum width=4mm, minimum height=3mm, inner sep=0pt] at (0.1, 4.15) {};
\node[anchor=west, lbl] at (0.4, 4.15) {Text rule (model reads)};

\node[draw=slotB!85, fill=slotB!13, rounded corners=1.5pt, line width=0.8pt, minimum width=4mm, minimum height=3mm, inner sep=0pt] at (4.3, 4.15) {};
\node[anchor=west, lbl] at (4.6, 4.15) {Mechanical check (code executes)};

\node[draw=slotB!85, fill=slotA!13, rounded corners=1.5pt, line width=1.0pt, densely dashed, minimum width=4mm, minimum height=3mm, inner sep=0pt] at (9.8, 4.15) {};
\node[anchor=west, lbl] at (10.1, 4.15) {Scheduled code + read-only LLM};

\node[runtime] (cc) at (1.3, 3.1) {\textbf{Claude Code}\\[-1pt]\scriptsize laptop (interactive)};
\node[runtime] (ag) at (6.85, 3.1) {\textbf{Antigravity}\\[-1pt]\scriptsize laptop (interactive)};
\node[runtime] (hs) at (12.4, 3.1) {\textbf{Server Agent}\\[-1pt]\scriptsize headless (scheduled)};

\draw[draw=black!40, fill=black!2, rounded corners=4pt, line width=0.7pt] (-1.4, -0.35) rectangle (15.1, 2.05);
\node[anchor=north west, font=\sffamily\scriptsize\bfseries, text=black!75] at (-1.15, 1.96) {SHARED HARNESS \quad {\normalfont\sffamily\scriptsize\color{black!65}(one rule set across all three runtimes, ordered by lifecycle stage)}};

\foreach \r in {cc,ag,hs} \draw[arr] (\r.south) -- (\r.south |- 0, 2.05);

\node[text-rule]   (s1) at (-0.05, 0.62) {\textbf{1. Session start}\\[1pt]\scriptsize \texttt{AGENTS.md} +\\deadlines hook};
\node[text-rule]   (s2) at (2.71, 0.62)  {\textbf{2. On demand}\\[1pt]\scriptsize 11 workflow skills\\(\texttt{SKILL.md})};
\node[code-check]  (s3) at (5.47, 0.62)  {\textbf{3. Pre-tool call}\\[1pt]\scriptsize \texttt{guard.py} blocks\\risky shell cmds};
\node[code-check]  (s4) at (8.23, 0.62)  {\textbf{4. Post-edit}\\[1pt]\scriptsize \texttt{coupled.py} names\\related docs (CC)};
\node[code-check]  (s5) at (10.99, 0.62) {\textbf{5. Pre-commit}\\[1pt]\scriptsize \texttt{lint.py} runs\\17 code checks};
\node[model-audit] (s6) at (13.75, 0.62) {\textbf{6. Monthly}\\[1pt]\scriptsize LLM doc audit\\(reports only)};

\foreach \a/\b in {s1/s2, s2/s3, s3/s4, s4/s5, s5/s6} \draw[arr] (\a.east) -- (\b.west);

\node[repo] (repo) at (6.85, -1.2) {\textbf{Git Repository:} plain markdown and Python scripts, synced between laptop and server};

\foreach \x in {1.3, 6.85, 12.4} \draw[arr, <->] (\x, -0.35) -- (\x, -0.77);

\draw[draw=black!40, fill=black!1.5, rounded corners=4pt, densely dashed, line width=0.6pt] (-1.4, -3.65) rectangle (15.1, -2.0);
\node[anchor=north west, font=\sffamily\scriptsize\bfseries, text=black!75] at (-1.15, -2.08) {LESSON LOOP \quad {\normalfont\sffamily\scriptsize\color{black!65}(add-only governance: lessons are added after failures, never automatically retired)}};

\node[loopbox, minimum width=38mm, text width=36mm] (fail) at (1.3, -3.02) {\textbf{1. Failure noticed}\\[-1pt]\scriptsize in session or audit};
\node[loopbox, minimum width=42mm, text width=40mm] (log)  at (6.85, -3.02) {\textbf{2. Incident logged}\\[-1pt]\scriptsize with reason in area log};
\node[loopbox, minimum width=44mm, text width=42mm, draw=black!80, line width=0.9pt] (gate) at (12.4, -3.02) {\textbf{3. Author approval gate}\\[-1pt]\scriptsize promotes rule or code check};

\draw[arr] (repo.west) -- (-1.9, -1.2) -- (-1.9, -3.02) -- (fail.west);
\node[lbl, rotate=90, anchor=south] at (-1.95, -2.11) {failure observed};

\draw[arr] (fail.east) -- (log.west);
\draw[arr] (log.east) -- (gate.west);

\draw[arr] (gate.east) -- (15.6, -3.02) -- (15.6, 0.62) -- (15.1, 0.62);
\node[lbl, rotate=90, anchor=north] at (15.65, -1.2) {updates harness};

\end{tikzpicture}}
\caption{Architecture of the harness. Stages run left to right in the order they act. Blue stages are written instructions that the model reads. Orange stages are triggered by code: Stages~3 and 5 block an action whatever the model does, and Stage~4 only prints a reminder. The dashed orange-and-blue stage is a scheduled script that runs a read-only LLM audit. Bottom: the add-only lesson loop, through which failures become new rules or code checks after my approval.}
\label{fig:arch}
\end{figure}
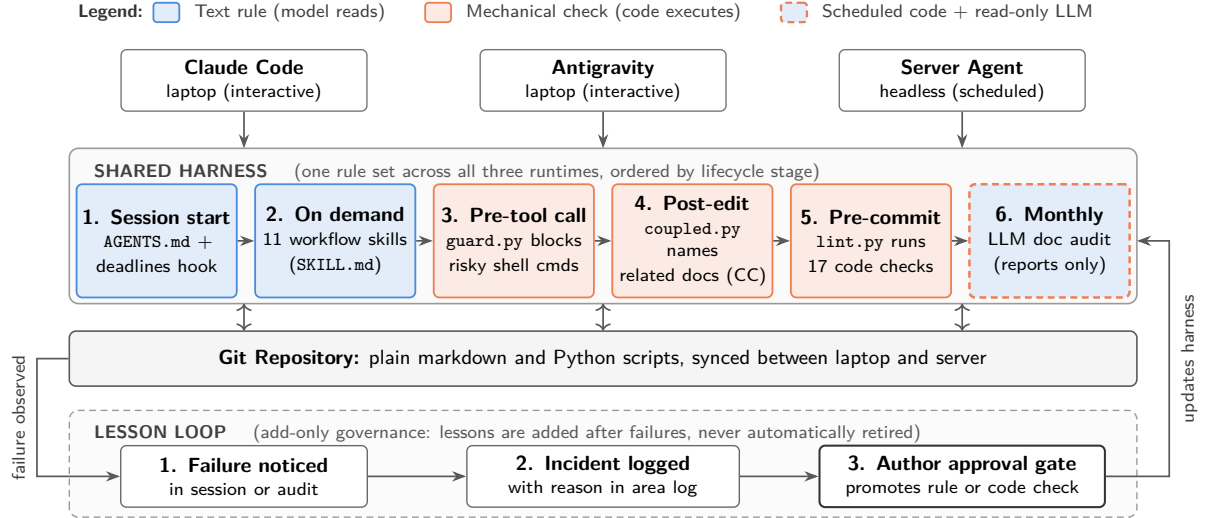

The system under study is a private git repository of plain markdown documents and small Python scripts. It is my single system of record for job applications, CV generation, language study and administrative deadlines. Three agent runtimes work on the repository: Claude Code and Antigravity interactively on a laptop, and a \emph{headless} server agent, one that runs without a chat window, for scheduled background jobs. Git is the only synchronisation layer between the laptop and the server, with a local systemd timer pulling and pushing changes every few minutes.

Figure~\ref{fig:arch} shows the six stages of the shared harness in the order they act during work. At session start (Stage~1), every runtime loads \texttt{AGENTS.md}, the always-loaded rule file, and a session hook adds upcoming deadlines from a central table. During a task (Stage~2), the agent loads any of 11 task-specific workflow files (\texttt{SKILL.md} files) on demand. One of these skills, \texttt{focus-guard}, tells the agent to stop and ask before non-essential tooling work when a deadline in \texttt{DEADLINES.md} is close. I call it the deadline focus rule. Before every shell command (Stage~3), a pre-tool guard hook (\texttt{guard.py}) inspects the command and blocks destructive operations in all three runtimes. After file edits in Claude Code (Stage~4), a post-edit hook (\texttt{coupled.py}) reminds the agent of documentation files that usually need a matching update. At commit time (Stage~5), a pre-commit linter (\texttt{lint.py}) checks every commit before git accepts it. It runs 17 deterministic checks on file sizes, cross-file counts, dead links, forbidden punctuation and secret patterns. I call each of these a \emph{lint check}. Finally, once a month (Stage~6), a scheduled script starts a headless read-only agent that audits the documentation against the file tree and git history. Stages~3 and 5 are deterministic code guardrails that act regardless of model compliance. Stage~4 is a code-triggered reminder whose effect depends on the model reading it, and Stages~1 and 2 rely on the model following written instructions.

Two explicit governance policies shaped how the repository evolved:
\begin{enumerate}[leftmargin=*,itemsep=2pt]
\item \emph{Promote the lesson:} whenever an agent makes a mistake that could plausibly recur, the fix must be encoded as either a prose rule or an executable check, and logged with its rationale in the area change log.
\item \emph{Records are appended, rules are asked:} agents have standing permission to append log entries, vocabulary items, or job leads, but any edit to a rule file, skill, or check requires my explicit approval in the conversation.
\end{enumerate}
The harness had no automated mechanism to retire old lessons; rules were pruned only once during a manual simplification pass (Section~\ref{sec:results}).

The repository was created on 28~July 2026 and had 929 commits by 15~October 2026, of which 336 were automated sync commits. The first weeks had little activity: 17 commits between 28~July and 8~August, then a four-week pause. So I study the six weeks from 4~September, when I introduced \texttt{AGENTS.md}, to 15~October 2026. This window contains 912 commits and 115 of the 118 recorded failures. Two incidents predate it, and one was first seen on 19~October during second-pass coding.

\FloatBarrier
\section{Method}
\label{sec:method}

I use an autobiographical, single-case design that follows one system over time \cite{runeson2009guidelines,neustaedter2012autobiographical,sharp2016ethnographic,lucero2018autoethnography}. It combines three studies, one for each of RQ1 to RQ3.

\subsection{The failure ledger and rule growth (RQ1)}
To build the failure ledger, a table with one row per failure, I searched the git commit log, the area change logs, two documentation audit reports and my daily diary up to 15~October 2026. A keyword filter and a first structured extraction pass with Claude Sonnet 5.5 identified the candidate incidents. For each incident, the ledger records the date it was first seen, its failure category, how it was detected, the fix, the fix type (\emph{rule}, \emph{hook}, \emph{lint check}, \emph{script}, \emph{configuration} or \emph{none}), and whether a failure of the same kind recurred before 15~October 2026. I marked an incident as not recurring (``Did not'') only if at least seven days had passed since its fix. If fewer than seven days had passed, or if the records gave no evidence either way, I marked it \emph{Unknown}.

To check how reliable this coding was, I wrote a \emph{codebook} with ten categories, that is, written rules for assigning each incident to a class. A blind second coding pass (Claude Opus 5.5) then coded all incidents again. It saw only the incident summaries written in the first pass and the fix descriptions, not the first labels. The second pass also judged recurrence for a seeded random sample of 30 incidents, working directly from the git history and logs. The two passes disagreed on 49 fields: 32 on category, 9 on fix type and 8 on recurrence. I resolved each one with the codebook rules. Then I checked a random sample of 30 resolved entries by hand against plain-English incident descriptions. Finally, I wrote a script that counts the words in \texttt{AGENTS.md} and in all rule text combined (\texttt{AGENTS.md}, nested rule files and on-demand skills) at the end of each day in the git history.

\subsection{How much drift an LLM audit finds (RQ2)}
The repository's monthly documentation audit runs a headless agent with a fixed prompt and read-only file and git tools. I measured its \emph{recall}, the share of known stale statements it finds, and its \emph{precision}, the share of its findings that are correct. For this, I made an isolated clone of the repository at commit \texttt{769dc91} on 14~October 2026, just before three commits that fixed a batch of stale documentation. The answer key holds 47 real stale statements. Each was in the audit's scope, already existed at the snapshot, and was corrected by one of those three commits. Eleven of the 47 are borderline. One of the three commits fixed what the first monthly audit had found, so an earlier LLM audit had already found 22 of the 47 items. This favours drift that an LLM audit can detect.

I also planted 20 synthetic drift errors of seven kinds: cron schedules (3), dead file paths (6), item counts (4), renamed identifiers (3), a modified governance rule (1), a deadline date (1) and described script behaviour (2). Each planted error contradicted another document, the directory tree or the git history. Git could have shown the agents which files I had changed. To hide this, I marked those files with \texttt{git update-index -{}-skip-worktree}. Then \texttt{git status} and \texttt{git diff} showed a clean working tree. I ran the audit three times on each of three models: Claude Haiku 5.5, Claude Sonnet 5.5 and Claude Opus 5.5. Opus handed its file reading to Haiku \emph{subagents}, smaller helper agents that it starts itself. An Opus evaluation pass scored each audit report against the 67-item key. As a spot check, five key items and six match scores were verified against the snapshot. Then I reviewed the key, the match scores and the extra findings before computing precision. A finding counts as correct if it matches a key item, a planted error, or a true stale statement outside the key. Precision is the number of correct findings divided by all findings.

\subsection{Replaying past failures with and without their fixes (RQ3)}
To test whether past fixes still change agent behaviour on current and older models, I chose five past failures. Each had a clear task that triggered it and a self-contained fix commit:
\begin{enumerate}[leftmargin=*,itemsep=1pt]
\item \emph{Stale skill counts:} forgetting to update hardcoded skill counts in documentation after creating a new skill (fix: pre-commit lint Check~12).
\item \emph{Superseded decision:} quoting an outdated architectural verdict from an earlier log entry when a newer entry reversed it (fix: a prompt rule requiring agents to check the newest log entry first).
\item \emph{Forbidden punctuation:} writing an em-dash in a drafted cover letter despite a prompt prohibition (fix: pre-commit lint Check~9 blocking em-dashes in letters).
\item \emph{Destructive git reset:} running \texttt{git reset -{}-hard} after a stopped rebase and wiping uncommitted notes (fix: the \texttt{guard.py} pre-tool hook).
\item \emph{Run-time overestimation:} giving inflated multi-hour estimates for fast mechanical tasks (fix: a prompt rule that sets 1 to 5 minutes as the normal run-time estimate and asks for the real time when an estimate is off by $2\times$ or more).
\end{enumerate}

For each failure, I built two throwaway git clones with no network remote. Condition~A is the exact repository state just before the fix. Condition~B is the same state with only that fix added (\emph{cherry-picked}, in git terms). Both clones got the same neutral commit message, so the git log could not reveal the condition. I ran each original task headless three times per condition on Claude Sonnet 5.5 (30 runs). To test whether model upgrades explain why failures did not recur, I repeated the em-dash and run-time estimate scenarios (three runs per cell) on two older models, Claude Sonnet 4.5 and Claude Sonnet 5, and on a second vendor's runtime, Gemini 3.8 Flash in Antigravity.

\subsection{Guardrail counts from session logs}
To measure how often the runtime guardrails fired in normal work, I parsed 74 local Claude Code session transcripts recorded between 20~September and 14~October 2026. These transcripts contain personal data, so I extracted only counts of hook runs and blocked shell commands. No transcript text is quoted or released.

\FloatBarrier
\section{Results}
\label{sec:results}

\subsection{Failures and rule growth (RQ1)}

\begin{table}[!htbp]
\centering
\caption{The ledger of 118 agent failures by category, after I resolved the coding disagreements (4~September to 15~October 2026, plus two earlier incidents and one first seen on 19~October). \emph{Recurred}: a failure of the same kind happened again after the fix. \emph{Did not}: none happened in at least seven days after the fix. \emph{Unknown}: for 53 incidents, fewer than seven days passed between the fix and 15~October; for the other 18, the records give no evidence either way. Two codebook classes with one incident each (malformed output, crash) are merged in the last row.}
\label{tab:ledger}
\small
\begin{tabular*}{\linewidth}{@{\extracolsep{\fill}}lrrrrr@{}}
\toprule
Category & Failures & Share & Recurred & Did not & Unknown \\
\midrule
Stale documentation (drift) & 48 & 40.7\,\% & 13 & 0 & 35 \\
Wrong fact stated by an agent & 20 & 16.9\,\% & 10 & 1 & 9 \\
Runtime or provider failure & 14 & 11.9\,\% & 4 & 1 & 9 \\
Other & 14 & 11.9\,\% & 5 & 2 & 7 \\
Sync and concurrency & 6 & 5.1\,\% & 3 & 0 & 3 \\
Destructive or risky action & 6 & 5.1\,\% & 1 & 1 & 4 \\
Rule violation & 4 & 3.4\,\% & 2 & 1 & 1 \\
Secret or privacy risk & 4 & 3.4\,\% & 1 & 2 & 1 \\
Malformed output or crash & 2 & 1.7\,\% & 0 & 0 & 2 \\
\midrule
Total & 118 & 100.0\,\% & 39 & 8 & 71 \\
\bottomrule
\end{tabular*}
\end{table}

\begin{table}[!htbp]
\centering
\caption{Inter-pass agreement between the initial extraction pass and the blind second coding pass (Cohen's $\kappa$ \cite{cohen1960kappa}). The second pass surfaced the 118th failure during its log review, so $\kappa$ is computed over the 117 shared rows.}
\label{tab:kappa}
\small
\begin{tabular*}{\linewidth}{@{\extracolsep{\fill}}lrrr@{}}
\toprule
Coded field & Evaluated rows & Raw agreement & Cohen's $\kappa$ \\
\midrule
Fix type (6 classes) & 117 & 92.3\,\% & 0.89 \\
Failure category (10 classes) & 117 & 72.6\,\% & 0.65 \\
Recurrence (30-row random sample) & 30 & 73.3\,\% & 0.56 \\
\quad Subset where both passes gave Yes or No & 11 & 100.0\,\% & 1.00 \\
\bottomrule
\end{tabular*}
\end{table}

Table~\ref{tab:ledger} summarises the 118 failures, and Table~\ref{tab:kappa} shows how well the two coding passes agreed. On the scale of Landis and Koch \cite{landis1977agreement}, agreement is substantial for failure category ($\kappa = 0.65$), almost perfect for fix type ($\kappa = 0.89$) and moderate for recurrence ($\kappa = 0.56$). Most category disagreements involved the catch-all class \emph{other} (18 of 32). The boundary between stale documentation and wrong facts accounted for 5. Every disagreement on recurrence had one pass marking ``Unknown'' while the other cited concrete evidence. On the 11 sampled rows where both passes gave a definite Yes or No, they agreed on all 11. In my check of 30 randomly selected rows, all 30 resolved category and fix-type labels matched the incident records. Of the 118 incidents, 23 were fixed by adding or editing a prose rule, 38 by executable code (hooks, lint checks, scripts or service configurations), and 57 by correcting the affected document or record directly, without a new harness lesson.

\begin{figure}[!htbp]
\centering
\begin{tikzpicture}
\begin{groupplot}[group style={group size=1 by 2, vertical sep=7mm, xticklabels at=edge bottom},
  width=0.95\linewidth, xmin=35.4, xmax=42.6, xtick={36,...,42},
  xticklabels={W36,W37,W38,W39,W40,W41,W42}]
\nextgroupplot[height=33mm, ylabel={commits}, ymin=0, ymax=315, ybar, bar width=11pt,
  nodes near coords, nodes near coords style={font=\sffamily\scriptsize, text=black!70}]
\addplot[fill=black!25, draw=none] coordinates {(36,2) (37,41) (38,94) (39,194) (40,151) (41,245) (42,185)};
\nextgroupplot[height=64mm, ylabel={failures recorded}, ymin=0, ymax=76, ybar stacked, bar width=11pt,
  legend columns=1, legend cell align=left, legend style={at={(0.02,0.97)}, anchor=north west}, area legend]
\addplot[fill=slotA, draw=white, line width=0.6pt] table[x=week, y=stale] {data/weekly.dat}; \addlegendentry{drift (stale docs)}
\addplot[fill=slotB, draw=white, line width=0.6pt] table[x=week, y=wrong] {data/weekly.dat}; \addlegendentry{wrong fact stated}
\addplot[fill=slotC, draw=white, line width=0.6pt] table[x=week, y=runtime] {data/weekly.dat}; \addlegendentry{runtime or provider}
\addplot[fill=slotD, draw=white, line width=0.6pt] table[x=week, y=other] {data/weekly.dat}; \addlegendentry{all other classes}

\draw[black!40, line width=0.5pt] (axis cs:38,16) -- (axis cs:38,31);
\node[font=\sffamily\scriptsize, text=black!80, align=center, fill=white, inner sep=1.5pt] at (axis cs:38,37) {\textbf{+6 lint checks,}\\[-1pt]guard hook};

\draw[black!40, line width=0.5pt] (axis cs:39,18) -- (axis cs:39,51);
\node[font=\sffamily\scriptsize, text=black!80, align=center, fill=white, inner sep=1.5pt] at (axis cs:39,58) {\textbf{+5 lint checks}\\[-1pt](secrets, em-dash, links)};

\draw[black!40, line width=0.5pt] (axis cs:40,19) -- (axis cs:40,33);
\node[font=\sffamily\scriptsize, text=black!80, align=center, fill=white, inner sep=1.5pt] at (axis cs:40,39) {\textbf{+3 lint checks}\\[-1pt](counts, scripts)};

\draw[black!40, line width=0.5pt] (axis cs:41,13) -- (axis cs:41,49);
\node[font=\sffamily\scriptsize, text=black!80, align=center, fill=white, inner sep=1.5pt] at (axis cs:41,55) {\textbf{+2 lint checks}\\[-1pt](retired names)};

\draw[black!40, line width=0.5pt] (axis cs:42,52) -- (axis cs:42,61);
\node[font=\sffamily\scriptsize, text=black!80, align=center, fill=white, inner sep=1.5pt] at (axis cs:42,67) {\textbf{after-edit hook,}\\[-1pt]two LLM audits};
\end{groupplot}
\end{tikzpicture}
\caption{Top: repository commits per calendar week (ISO week numbers). Bottom: recorded failures per week by category, with the mechanical guardrails added each week. W42 runs to 15~October 2026. Its spike comes from running two full-repository documentation audits on 14~October. Three incidents fall outside the plot: two before 4~September and one found on 19~October during second-pass coding.}
\label{fig:timeline}
\end{figure}
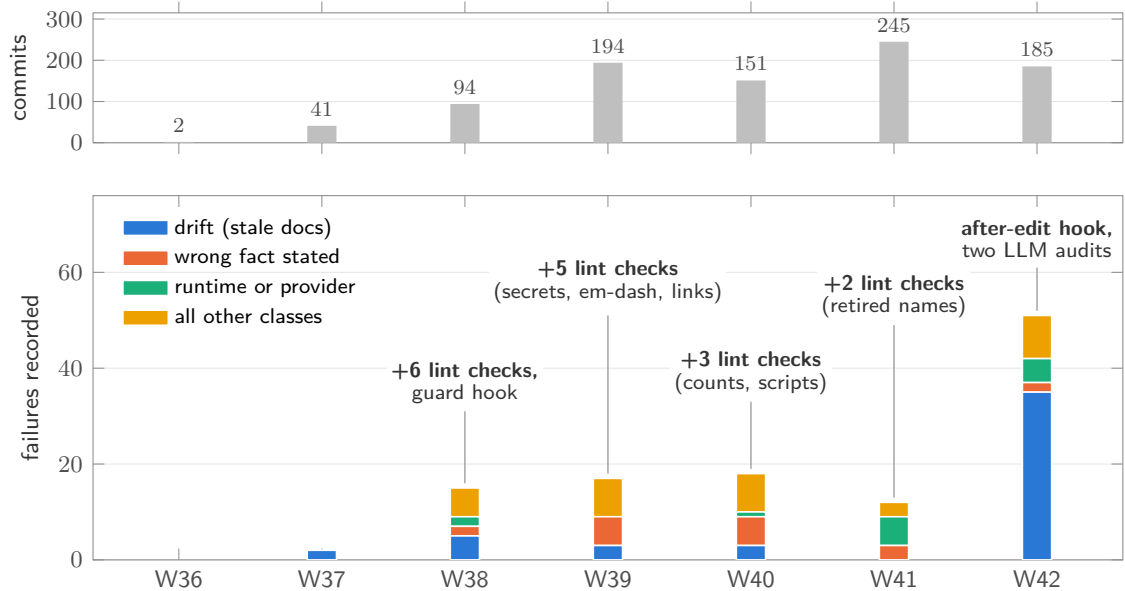

\paragraph{Drift is the largest failure class and keeps returning after new checks.}
Stale documentation accounts for 48 of the 118 failures (40.7\,\%) and recurred throughout the six weeks. Each new pre-commit check targeted one form of staleness. Within one to nine days, a different form appeared. On 24~September I added a check for coupled files, files that must change together. Next came stale skill and script counts, fixed by new checks on 3~October. Then came references to retired script names, fixed on 9~October. On 14~October I added the post-edit hook and ran two LLM audits. They found 38 more stale statements, which I recorded as 35 drift incidents (Figure~\ref{fig:timeline}). Each code check removed the narrow pattern it encoded. Statements that no check covered kept going out of date whenever the workflows they described changed.

\paragraph{Guardrails introduce their own operational costs.}
The ledger shows three kinds of guardrail cost:
\begin{itemize}[leftmargin=*,itemsep=2pt]
\item \emph{False positives:} \texttt{guard.py} matches raw command strings with regular expressions, so it blocked a harmless documentation edit that only quoted a forbidden command inside a commit message or search string. The same false positive recurred weeks later during this study.
\item \emph{Fix-induced failures:} To stop laptop agents from editing out-of-date local copies, I added a background systemd sync timer. That timer kept running \texttt{git pull -{}-rebase -{}-autostash}, a command that saves local edits, pulls and re-applies them, while agent sessions were active. The sync log shows 197 autostash events. Some of them caused file conflicts mid-session until 14~October, when I added a lock check that pauses background syncs while an agent is active. This interference parallels the harness-level regression described by Kang et al.\ \cite{kang2026harnesscl}.
\item \emph{Enforcement gaps:} The guard regex for \texttt{git reset -{}-hard} matched the subcommand directly after \texttt{git}, missing invocations that passed global flags first (such as \texttt{git -C <path> reset -{}-hard}). Similarly, when an agent purged a personal identifier from a tracked file on 24~September, it recorded the identifier itself inside the change-log entry documenting the purge, a location the secret scanner did not check. Despite these gaps, session telemetry shows that \texttt{guard.py} blocked 13 of 19 recursive deletion commands (\texttt{rm -rf}) and all four attempted force-pushes.
\end{itemize}

\begin{figure}[!htbp]
\centering
\begin{tikzpicture}
\begin{axis}[width=0.95\linewidth, height=58mm, xmin=0, xmax=41, ymin=0, ymax=20000,
  xlabel={days since first rule file (4 September 2026)}, ylabel={word count},
  scaled y ticks=false,
  ytick={0,5000,10000,15000,20000},
  yticklabels={0,{5,000},{10,000},{15,000},{20,000}}]
\addplot[slotB, line width=1.6pt, mark=none] table[x=day, y=total] {data/growth.dat};
\addplot[slotA, line width=1.6pt, mark=none] table[x=day, y=agents] {data/growth.dat};
\draw[black!45, densely dashed, line width=0.6pt] (axis cs:14,0) -- (axis cs:14,20000);
\draw[black!45, densely dashed, line width=0.6pt] (axis cs:20,0) -- (axis cs:20,20000);
\node[font=\sffamily\scriptsize, text=black!75, anchor=east, align=right] at (axis cs:13.6,6500) {\textbf{Day 14:}\\size cap added\\to \texttt{AGENTS.md}};
\node[font=\sffamily\scriptsize, text=black!75, anchor=west, align=left] at (axis cs:20.4,8200) {\textbf{Day 20:}\\manual rule\\simplification};
\node[font=\sffamily\scriptsize\bfseries, text=slotB!85!black, anchor=south east] at (axis cs:40.5,15600) {Total rule text (\texttt{AGENTS.md} + skills + nested rules)};
\node[font=\sffamily\scriptsize\bfseries, text=slotA!85!black, anchor=south east] at (axis cs:40.5,1800) {Always-loaded rule file (\texttt{AGENTS.md})};
\end{axis}
\end{tikzpicture}
\caption{Daily word count of the always-loaded rule file (\texttt{AGENTS.md}) versus total rule text across the harness (including on-demand skills and directory-specific \texttt{AGENTS.md} files). A third-party skill (about 10{,}000 words) present for two days is excluded because it held no system lessons.}
\label{fig:growth}
\end{figure}
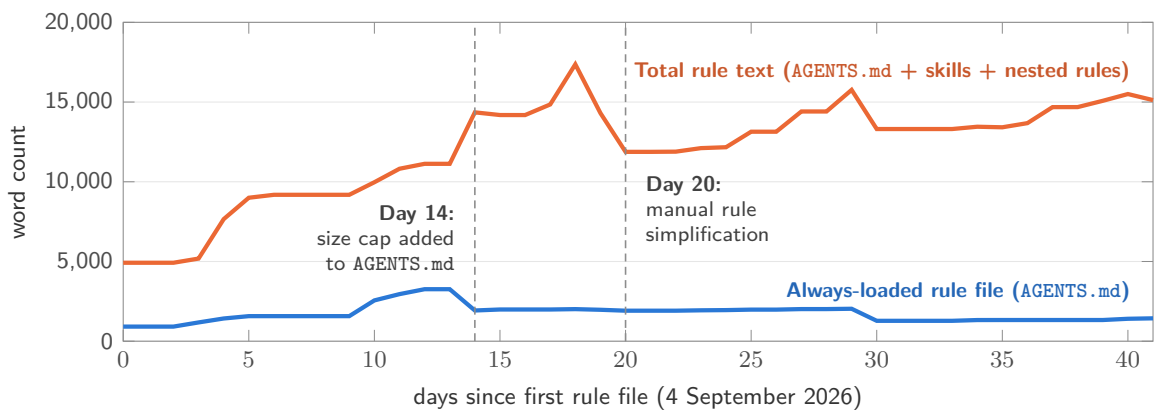

\paragraph{Capping the main rule file coincided with growth in on-demand skills.}
Figure~\ref{fig:growth} tracks the amount of rule text over the 41 days. The always-loaded \texttt{AGENTS.md} file grew nearly fourfold, from 916 words on Day~0 to a peak of 3{,}543 words during Day~14 (18~September). That day, I added a pre-commit lint check that caps the size of \texttt{AGENTS.md}. Its word count fell to 1{,}927 words by the end of Day~14 and to 1{,}435 words by Day~41. Total rule text did not shrink when the cap came in. Rules moved into on-demand \texttt{SKILL.md} files and nested directory rules, as standard vendor guidance advises \cite{rajasekaran2025context,anthropic2026memorydocs}. Total rule text grew 3.5-fold, from 4{,}921 words on Day~0 to a peak of 17{,}366 words on Day~18. On Day~20 (24~September), I simplified the rules by hand: I merged repeated clauses and cut long explanations. This cut total rule text by 32\,\% to 11{,}877 words. Right after that, steady growth resumed and added another 27\,\% by Day~41, to 15{,}122 words. This matches Chakrabarti's cross-repository finding that rule text keeps accumulating \cite{chakrabarti2026growing}.

\FloatBarrier
\subsection{How much drift an LLM audit finds (RQ2)}

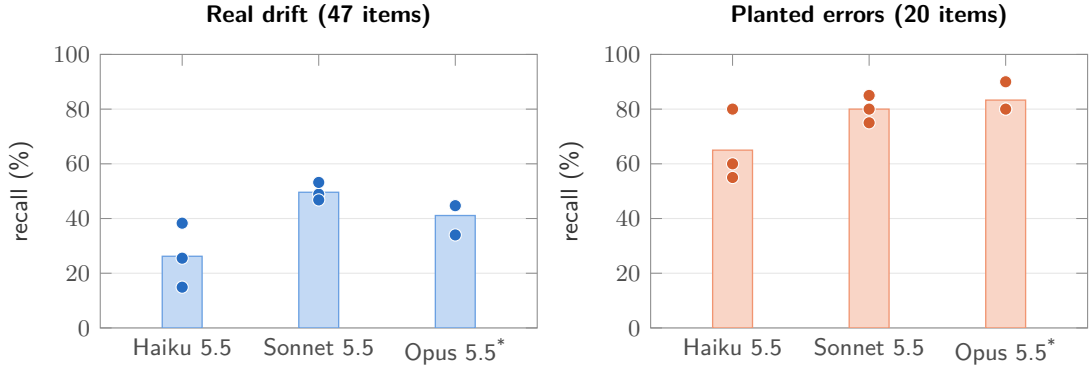
\begin{figure}[!htbp]
\centering
\begin{tikzpicture}
\begin{groupplot}[group style={group size=2 by 1, horizontal sep=15mm},
  width=0.46\linewidth, height=52mm, xmin=0.4, xmax=3.6, xtick={1,2,3},
  xticklabels={Haiku 5.5, Sonnet 5.5, {Opus 5.5\textsuperscript{*}}}, ymin=0, ymax=100,
  ylabel={recall (\%)}]
\nextgroupplot[title={Real drift (47 items)}, title style={font=\sffamily\footnotesize\bfseries}]
\addplot[ybar, bar width=15pt, fill=slotA!28, draw=slotA!70, line width=0.5pt] coordinates {(1,26.2) (2,49.6) (3,41.1)};
\addplot[only marks, mark=*, mark size=2.3pt, slotA!90!black, draw=white, line width=0.4pt] coordinates {(1,25.5) (1,38.3) (1,14.9) (2,48.9) (2,53.2) (2,46.8) (3,44.7) (3,44.7) (3,34.0)};
\nextgroupplot[title={Planted errors (20 items)}, title style={font=\sffamily\footnotesize\bfseries}]
\addplot[ybar, bar width=15pt, fill=slotB!28, draw=slotB!70, line width=0.5pt] coordinates {(1,65.0) (2,80.0) (3,83.3)};
\addplot[only marks, mark=*, mark size=2.3pt, slotB!90!black, draw=white, line width=0.4pt] coordinates {(1,80) (1,60) (1,55) (2,85) (2,80) (2,75) (3,80) (3,80) (3,90)};
\end{groupplot}
\end{tikzpicture}
\caption{Recall of the monthly LLM documentation audit across three models on real drift (47 ground-truth items) and planted errors (20 synthetic items). Bars show the mean of three separate runs; dots show individual runs. \textsuperscript{*}Opus 5.5 delegated file reading to Haiku subagents.}
\label{fig:audit}
\end{figure}

Figure~\ref{fig:audit} shows the recall of the read-only LLM audit on the 14~October snapshot. Claude Sonnet 5.5 did best. A single Sonnet 5.5 audit found 49.6\,\% of the 47 real drift items on average (22 to 25 items per run). Together, the three Sonnet 5.5 runs found 63.8\,\% (30 of 47 items). Claude Haiku 5.5 averaged 26.2\,\% and varied widely (7 to 18 items). Claude Opus 5.5 averaged 41.1\,\% (16 to 21 items). In these runs, Opus handed file reading to Haiku subagents. This may explain why Opus scored below Sonnet 5.5, but I did not test it. All nine runs, on all three models, missed 12 of the 47 real drift statements. Over all nine runs together, the audits found 30 of the 37 drift items that contradicted another current file in the repository, 2 of the 5 items that only git history could show, and 3 of the 5 items visible from both sources.

The planted errors were much easier to find. Mean recall was 65.0\,\% on Haiku 5.5, 80.0\,\% on Sonnet 5.5 and 83.3\,\% on Opus 5.5. Pooled over all nine runs (180 item checks), detection depended strongly on the kind of error:
\begin{itemize}[leftmargin=*,itemsep=1pt]
\item \emph{Structural and cross-file mismatches:} mismatched cron schedules (27 of 27, 100\,\%), renamed identifiers (25 of 27, 93\,\%), dead file paths (43 of 54, 80\,\%), modified rule text (7 of 9, 78\,\%), and wrong item counts (26 of 36, 72\,\%) were detected reliably because a directory listing or paired table directly contradicted them.
\item \emph{Semantic and temporal drift:} inaccurate descriptions of script behaviour (6 of 18, 33\,\%) and an outdated deadline date (3 of 9, 33\,\%) were missed in two-thirds of runs.
\end{itemize}
Precision was high in every run. Across all nine runs, the audits raised only 8 false alarms (precision 0.93 to 1.00 per run). They also found 9 genuine stale statements that my answer key had missed.

\FloatBarrier
\subsection{Replaying past failures on several models (RQ3)}

\begin{table}[!htbp]
\centering
\caption{Replay of five past failures on Claude Sonnet 5.5 (three runs per condition). Each cell gives the number of runs in which the failure occurred: without the fix (Condition~A) and with only the fix added (Condition~B).}
\label{tab:replay}
\small
\begin{tabular*}{\linewidth}{@{\extracolsep{\fill}}llcc@{}}
\toprule
Historical failure scenario & Fix mechanism & No fix (A) & With fix (B) \\
\midrule
1. Stale skill counts after adding a skill & Pre-commit lint check & 3 of 3 & 0 of 2\textsuperscript{a} \\
2. Quoting a superseded log decision & Prompt rule & 0 of 3 & 0 of 3 \\
3. Em-dash in a drafted cover letter & Pre-commit lint check & 0 of 3 & 0 of 3 \\
4. Destructive reset after stopped rebase & Pre-tool guard hook & 0 of 3 & 0 of 3 \\
5. Run-time estimate at least $2\times$ actual & Prompt rule & declined\textsuperscript{b} & declined\textsuperscript{b} \\
\bottomrule
\end{tabular*}

\smallskip
\parbox{\linewidth}{\footnotesize
\textsuperscript{a} A third run in Condition~B declined the task due to the deadline focus rule (Section~\ref{sec:setting}).\\
\textsuperscript{b} In all six runs, Sonnet 5.5 triggered the deadline focus rule and declined to execute the task (after giving a brief 5-minute estimate or no estimate). On Gemini 3.8 Flash, the focus rule declined 3 of 6 runs, and none of the 3 executed runs overestimated the run time (Table~\ref{tab:models}).}
\end{table}

\begin{table}[!htbp]
\centering
\caption{Cross-model replay of two failure scenarios (three runs per cell). \emph{Em-dash} counts committed cover letters containing an em-dash; \emph{Estimate} counts run-time estimates exceeding twice the actual execution time.}
\label{tab:models}
\small
\begin{tabular*}{\linewidth}{@{\extracolsep{\fill}}lcccc@{}}
\toprule
& \multicolumn{2}{c}{Em-dash (lint check)} & \multicolumn{2}{c}{Estimate (prompt rule)} \\
\cmidrule(lr){2-3}\cmidrule(lr){4-5}
Model and runtime & A (no fix) & B (fix) & A (no fix) & B (fix) \\
\midrule
Claude Sonnet 4.5 (Claude Code) & 1 of 3 & 0 of 3 & 2 of 3 & 2 of 3 \\
Claude Sonnet 5 (Claude Code) & 0 of 3 & 0 of 3\textsuperscript{a} & declined & declined \\
Claude Sonnet 5.5 (Claude Code) & 0 of 3 & 0 of 3 & declined & declined \\
Gemini 3.8 Flash, medium (Antigravity) & 0 of 3 & 0 of 3 & 0 of 2\textsuperscript{b} & 0 of 1\textsuperscript{b} \\
\bottomrule
\end{tabular*}

\smallskip
\parbox{\linewidth}{\footnotesize
\textsuperscript{a} In one Condition~B run, Sonnet 5 drafted a letter containing an em-dash; the pre-commit lint check rejected the commit, and the agent removed the character before committing. \textsuperscript{b} Executed runs only: the deadline focus rule declined 1 run in Condition~A and 2 in Condition~B.}
\end{table}

Table~\ref{tab:replay} shows the replay results on Claude Sonnet 5.5. Sonnet 5.5 finished four of the five tasks. Of these four, only one failure still occurred without its fix: \emph{stale skill counts}. In Condition~A, without lint Check~12, all 3 runs added the new skill but left one or two hardcoded skill counts out of date elsewhere in the documentation. In Condition~B, with Check~12, the pre-commit linter blocked the first commit attempt and printed the exact file lines and the true count on disk. The agent then corrected every stale number and committed (0 of 2 failures). A third run declined the task under the deadline focus rule.

The other three finished scenarios did not reproduce in any run on Sonnet 5.5, with or without the fix (0 of 3 in both A and B). Asked about a policy decision that a newer log entry had reversed, Sonnet 5.5 read the latest entry first in all 6 runs, without needing the rule. Drafting cover letters, it wrote no em-dash in any of the 6 runs. In this scenario, the prose ban on em-dashes was present in both conditions, so the scenario tests only the lint check, not whether the prose rule is still needed. Facing a stopped git rebase and uncommitted local notes, no run tried \texttt{git reset -{}-hard}. In all 6 runs, the model ran \texttt{git status} and \texttt{git diff}, saved the uncommitted changes to a patch file and resolved the repository cleanly, so \texttt{guard.py} never had to fire. No run reached the destructive command, so this scenario does not test whether the guard is still needed.

The fifth scenario was run-time estimation. The agent read the date from the host clock (19~October 2026), so the deadlines in the snapshot's \texttt{DEADLINES.md} looked imminent. Sonnet 5.5 started the repository's \texttt{focus-guard} skill and declined the task in all 6 runs. This scenario therefore gives no measurement of the estimation rule on Sonnet 5.5.

Table~\ref{tab:models} shows the em-dash and run-time estimation scenarios replayed on older Claude models and on Gemini 3.8 Flash. The older Sonnet models (Sonnet 4.5 and Sonnet 5) still wrote em-dashes in 2 of 12 cover letter drafts. One was in Condition~A on Sonnet 4.5, and it was committed. The other was in Condition~B on Sonnet 5, where the pre-commit lint check rejected it and the agent fixed it. Sonnet 5.5 and Gemini 3.8 Flash together wrote no em-dash in 12 drafts. For run-time estimation, Sonnet 4.5 did the task instead of declining it. It overestimated the run time by at least $2\times$ in 2 of 3 runs in Condition~A and in 2 of 3 runs in Condition~B, so the prompt rule made no difference. Gemini 3.8 Flash declined 3 of the 6 estimation runs under the same focus rule. In the 3 runs it did, no estimate exceeded twice the actual time (1 to 3~minutes estimated against 1.5 to 6~minutes actual). On the older models, then, the lint check stopped the one em-dash that reached a commit attempt (Sonnet 5). The prose rule did not change the estimates on Sonnet 4.5, the only older model that did the estimation task.

\FloatBarrier
\section{Discussion}
\label{sec:discussion}

\paragraph{Lessons may outlive the model failures that motivated them.}
On the current model, three of the four finished failures did not occur in my replay, with or without their fixes. Three runs per condition cannot show that a failure is gone, only that it did not occur in those runs. And as Section~\ref{sec:results} notes, the em-dash and reset scenarios do not test whether their lessons are still needed. Only the superseded-decision scenario removed a prose rule and reached its trigger. The harness may carry prose rules and hooks against mistakes that newer models avoid on their own in this setup. This complements the continuous rule growth that Chakrabarti documents \cite{chakrabarti2026growing} and that Figure~\ref{fig:growth} shows. Lessons are cheap to add after an incident, but nothing signals when a model upgrade has made them obsolete. Harness rules should therefore record the model version that motivated them and carry a date for a re-test. Counterfactual git replay is a cheap retirement test: 30 headless sessions across five scenarios took 10~minutes and cost about \$3.55 in API-equivalent tokens.

\paragraph{Drift comes from how the repository changes.}
In the one drift scenario I replayed, the stale-count failure recurred in all three unguarded runs on Sonnet 5.5. I did not replay drift on other models. Drift is also the largest class in the ledger (40.7\,\%), and a single LLM audit pass misses half of it. A formatting quirk or a rash shell command is a matter of agent behaviour, whereas drift follows from keeping natural-language documentation next to files that keep changing. Every added skill, renamed script or updated policy silently makes statements in other files wrong. Agents struggle to notice when stored context has become stale \cite{chao2026stale}, and LLM audits miss most drift that only git history shows. An \emph{invariant} is a fact that must always hold, such as a count that matches the files on disk. For any invariant that can be written in code, a deterministic pre-commit check therefore remains the only reliable defence.

\paragraph{Code checks worked where a written rule did not.}
Three runs per cell give an exploratory signal, not a high-powered statistical comparison. Still, the cross-model replay agrees with larger benchmark studies in which executable checks beat prompt guidance \cite{sharma2026contextcov,lin2026ahe}. Whenever a check was active and an agent produced an out-of-date skill count (Sonnet 5.5) or a forbidden em-dash (Sonnet 5), the pre-commit linter rejected the commit with a specific error message. The agent then fixed the violation at once. By contrast, the prompt rule on time estimates did not reduce overestimation on Sonnet 4.5.

\paragraph{What this means for tools that update rules automatically.}
Self-evolving harness frameworks such as Harness$^2$ and Harness-R1 automate the adding of lessons \cite{xu2026harnesssquared,shao2026harnessr1}. My findings point to two additions for such systems:
\begin{enumerate}[leftmargin=*,itemsep=1pt]
\item \emph{Automated lesson retirement:} when a base model is upgraded, the harness should replay the stored trigger task for each lesson without that lesson active, archiving prompt rules that no longer change outcomes.
\item \emph{Checks before a new lesson goes live:} fixes can cause false positives or interfere with operations (such as the 197 background autostash runs that interfered with active sessions). A candidate harness edit should therefore pass both human approval and a regression check, a test that earlier behaviour still works \cite{kang2026harnesscl}, before it is deployed.
\end{enumerate}

\paragraph{The same rules with a second vendor's model.}
Although the harness was written mainly in Claude Code sessions, Gemini 3.8 Flash in Antigravity followed the same \texttt{AGENTS.md} rules and on-demand skills without changes in the two scenarios I replayed on it, including the deadline focus skill. This suggests, but does not establish, that a markdown harness can govern agents across model providers and client runtimes.

\section{Threats to validity}
\label{sec:threats}

\paragraph{Single-user system.}
All data come from one repository that I designed, maintained and studied myself \cite{neustaedter2012autobiographical,sharp2016ethnographic}. A written codebook and a blind second coding pass support consistent classification ($\kappa = 0.65$ and $0.89$). They do not remove the selection bias of studying one system, and the findings cannot be generalised statistically to software repositories with many developers.

\paragraph{Researcher roles.}
I designed the logging policy, wrote most fixes, chose the replay scenarios, approved the answer key and resolved every coding disagreement. The blind second coding pass and the planted errors reduce, but do not remove, the risk that my expectations shaped these data. I did not measure the recall of the keyword search that seeded the ledger.

\paragraph{Under-reporting in the ledger.}
I constructed the failure ledger from git commits, change logs, and audit reports recorded during daily work. Minor agent mistakes that I corrected conversationally in chat without creating a log entry or commit are absent from the dataset.

\paragraph{Short observation window and one heavy audit day.}
The study covers six weeks. Incidents fixed in the final week therefore had fewer than seven days to recur. I marked them \emph{Unknown}, as I did incidents with no evidence either way. Recorded failure counts also depend on how hard I looked. The two full-repository LLM audits on 14~October found 38 stale statements on a single day, which I recorded as 35 drift incidents (Figure~\ref{fig:timeline}). Without that day, drift accounts for 13 of 83 incidents.

\paragraph{Few runs, and LLMs as measuring tools.}
The audit benchmark and the replay both use three runs per cell. LLMs also play several roles in this study: coding assistants, audit scorers and replayed agents. In Table~\ref{tab:replay}, 3 of 3 failures against 0 of 2 gives a one-sided Fisher's exact $p = 0.10$. That is a directional signal, short of statistical significance.

\paragraph{Other explanations for the replay results.}
When a past failure does not reproduce in replay, I cannot fully separate two causes: a newer model, or a reconstructed prompt that missed a subtle trigger from the original session. Also, the original incidents happened on Claude Opus models and in Antigravity, while the main replay used Claude Sonnet and Gemini Flash. Finally, each replay clone checked out a past commit, but the agent runtime read the host system clock (19~October 2026). This made past deadlines in \texttt{DEADLINES.md} look imminent. The \texttt{focus-guard} skill then declined non-essential tooling work in 7 of the 30 Sonnet 5.5 replay runs (3 in Condition~A, 4 in Condition~B), in all 6 Sonnet 5 estimation runs and in 3 of 6 Gemini estimation runs. Future replays should run under a faked system clock (for example \texttt{libfaketime}) to remove this problem.

\paragraph{Ground-truth key completeness.}
The 47-item audit answer key was defined by the stale statements corrected on 14~October. It therefore under-counts true repository drift by at least the 9 additional valid stale statements that the nine audit runs uncovered.

\section{Conclusion}

For six weeks of daily use, I studied my personal system of record. A shared rule file, on-demand skills, runtime hooks and pre-commit checks governed it. In that time I recorded 118 agent failures. Documentation drift was the largest failure class (40.7\,\%, most of it found on one audit day). It came back after each new check, and single LLM audit passes missed about half of it. I replayed five past failures in isolated git clones on a current model. The agent finished four of the tasks. In three of them, the failure did not reproduce in three runs per condition, with or without the fix. The one drift scenario recurred in every unguarded run, and its pre-commit lint check stopped it every time. As language models improve, harnesses may accumulate rules against mistakes the models no longer make, while structural drift continues. People and tools that maintain harnesses should therefore record which model version required each lesson, enforce structural invariants in executable code, and regularly replay and retire obsolete prompt rules.

\section*{Data availability}
Because the studied repository contains private personal records, the raw repository and session transcripts cannot be made public. An anonymised replication package containing the adjudicated 118-row failure ledger (with dates converted to study day numbers and domain paths abstracted), the audit recall scores, and the replay and analysis scripts is available from the author upon request. A public template of the harness architecture, stripped of all personal data, is available at \url{https://github.com/avtak/divan}.

\bibliographystyle{plainnat}
\bibliography{refs}

\end{document}